\documentclass{article}
\usepackage{amsfonts}
\usepackage{amsmath}
\usepackage{amssymb}
\usepackage{amsthm}
\usepackage{mathrsfs}
\usepackage{bm}
\usepackage{enumerate}
\usepackage{hyperref}
\usepackage{authblk}
\usepackage[all]{xy}
\newcommand{\ud}{\mathrm{d}}
\newcommand{\dev}{\partial}
\newcommand{\R}{\mathbb{R}}
\newcommand{\Z}{\mathbb{Z}}
\newcommand{\A}{\mathcal{A}}

\newcommand{\X}{\mathfrak{X}}

\newcommand{\mdev}{\boldsymbol{\dev}}
\newcommand{\mlambda}{\boldsymbol{\lambda}}
\newcommand{\mmu}{\boldsymbol{\mu}}

\renewcommand{\L}{\mathcal{L}}

\newcommand{\<}{\langle}
\renewcommand{\>}{\rangle}
\renewcommand{\vec}[1]{\mathbf{#1}}
\newtheorem{thm}{Theorem}

\theoremstyle{definition}
\newtheorem{deff}{Definition}

\title{Dispersive deformations of the Hamiltonian structure of Euler's equations}
\author{M.~Casati}
\affil{Scuola Internazionale Superiore di Studi Avanzati\\via Bonomea 265, 34136 Trieste (ITALY)}

\begin{document}
\maketitle
\begin{abstract}
 Euler's equations for a two dimensional system can be written in Hamiltonian form, where the Poisson bracket is the Lie--Poisson bracket associated to the Lie algebra of divergence-free vector fields.  We show how to derive the Poisson brackets of 2d hydrodynamics of ideal fluids as a reduction from the one associated to the full algebra of vector fields. Motivated by some recent results about the deformations of Lie--Poisson brackets of vector fields, we study the dispersive deformations of the Poisson brackets of Euler's equation and show that, up to the second order, they are trivial.\end{abstract}
\section{Introduction}\label{sec:intro}
The idea that the dynamics of fluids can be understood using the language of infinite dimensional Lie groups is due to Arnol'd and dates back from the 1960's \cite{Ar66}. In that seminal paper, Arnol'd proved the correspondance between the motion of an ideal fluid and the geodesics of the right invariant metric corresponding to the kinetic energy of the fluid itself.

The main idea is that the configuration space of the system is the Lie group of the volume--preserving diffeomorphisms of the domain occupied by the fluid;the diffeomorphisms must be volume--preserving because an ideal fluid is incompressible.

In this paper, we will study the Euler's equations for a two-dimensional fluid. They can be put in Hamiltonian form, that essentially corresponds to Helmholtz's equation for the vorticity \cite{O82}. We will write
\begin{equation}\label{eq:HelmHam}
 \frac{\dev \omega}{\dev t}=\{H,\omega\}
\end{equation}
 where $\omega$ is the scalar vorticity of the fluid, $H$ is the Hamiltonian function corresponding to the kinetic energy and the Poisson brackets are the \emph{Lie--Poisson brackets} associated to the algebra of the divergence free vector fields. Such an algebra is exactly the Lie algebra of the group of the volume preserving diffeomorphisms introduced by Arnol'd.

We will show how to derive the bracket from the more general Lie--Poisson bracket of vector fields, introduced by Novikov \cite{N82} and corresponding to the Hamiltonian structure of the so-called EPDiff equation \cite{HM05}. Motivated by our study of the dispersive deformations of the Poisson brackets of hydrodynamic type \cite{MC15-1}, we study the same kind of deformations for the Poisson structure of Euler's equation. In doing so, we rely on the language of multidimensional Poisson Vertex Algebras (mPVA), that constitute a very efficient framework for the needed computations and in general for the study of Hamiltonian PDEs \cite{BdSK09}.

The well-known Euler's equations for a 2-dimensional ideal fluid are
\begin{equation}
\begin{cases}
 \frac{\dev \vec{u}}{\dev t}+\vec{u}\cdot\nabla \vec{u}+\nabla p=0\\
 \nabla\cdot\vec{u}=0
\end{cases},
\end{equation}
where $\vec{u}=(u,v)$ is the velocity field, $p$ is the pressure and $\nabla=\sum_{j=1}^2\vec{e}_j\frac{\dev}{\dev x^j}$.

By Helmholtz's decomposition theorem for vector fields, the two-dimensional ideal fluid is characterized by its vorticity alone
\begin{equation}
 \omega=\nabla\times\vec{u}=\dev_xv-\dev_yu.
\end{equation}
Taking the curl of Euler's equation we derive the so-called Helmholtz's equation \cite{L32}
\begin{equation}\label{eq:Helm}
 \omega_t+u\dev_x\omega+v\dev_y\omega=0
\end{equation}
which, as observed by Olver, can be written in Hamiltonian form \eqref{eq:HelmHam}. The Poisson brackets have form 
\begin{equation}\label{eq:PB-def}
 \{\vec{\omega}(\vec{r}),\vec{\omega}(\vec{r}')\}=\left(\dev_x \omega(\vec{r})\frac{\ud}{\ud y}-\dev_y\omega(\vec{r})\frac{\ud}{\ud x}\right)\delta(\vec{r}-\vec{r}')
\end{equation}
where we denote $\vec{r}=(x,y)$ and $\vec{r}'=(x',y')$.
To derive Equation \eqref{eq:Helm} from Equation \eqref{eq:HelmHam}, we recall that for two-dimensional fluids the velocity field can be uniquely determined by introducing the \emph{stream function} $\psi$, the scalar analogue of the vector potential for a solenoidal (or the soleinodal component according to Helmholtz's decomposition) field. By definition, we have
\begin{align}\label{eq:stream}
 u&=\frac{\dev \psi}{\dev y}&v&=-\frac{\dev\psi}{\dev x}.
\end{align}
The Hamiltonian $H$, hence, can be written as
\begin{equation}\label{eq:Ham-manip}
 \begin{split}
   H&=\frac{1}{2}\int|\vec{u}(\vec{r})|\ud\vec{r}=\frac{1}{2}\int\psi_x^2(\vec{r})+\psi_y^2(\vec{r})\ud\vec{r}\\&=-\frac{1}{2}\int\omega(\vec{r})\psi(\vec{r})\ud\vec{r}
 \end{split}
\end{equation}
where we integrated by parts to get the final result. By inserting in the expression for $H$ the solution of the Poisson equation $\nabla^2\psi=-\omega$, whose validity can be easily checked, we can write
\begin{equation}
 H=\frac{1}{2}\iint \omega(\vec{r}')\log |\vec{r}-\vec{r}'|\omega(\vec{r})\ud\vec{r}\ud\vec{r}'
\end{equation}
and hence, finally, we can find the variational derivative of $H$ with respect to $\omega$
\begin{equation}
 \frac{\delta H}{\delta \omega(\vec{s})}=\psi(\vec{s}).
\end{equation}
The computation of the Poisson bracket on the RHS of \eqref{eq:HelmHam} is done according to the usual general formula
\begin{equation}\label{eq:PoissVec}
\{F,G\}=\iint\frac{\delta F}{\delta \omega(\vec{r})}\{\omega(\vec{r}),\omega(\vec{r}')\}\frac{\delta G}{\delta \omega(\vec{r}')}\ud\vec{r}\ud\vec{r}',
\end{equation}
that for $F=H$ and $G=\omega(\vec{r}')$ gives $-u\omega_x-v\omega_y$.
\section{The Lie--Poisson bracket}
Let $\mathfrak{g}$ be a Lie algebra, and consider its dual space $\mathfrak{g}^*$. We can define a linear Poisson bracket between functions $F,G\in C^\infty(\mathfrak{g}^*)$, which is called the \emph{Lie--Poisson bracket}. The De Rham differential of functions on $\mathfrak{g}^*$ that we denote, in general, $\ud F(\alpha)$ ($\alpha\in\mathfrak{g}^*$), is a linear map $T_{\alpha}\mathfrak{g}^*\to\R$; since the tangent bundle of a vector space coincides with the underlying space, we can regard the differential as a linear map $\mathfrak{g}^*\to\R$, hence as an element of $\mathfrak{g}^{**}$, hence as an element of $\mathfrak{g}$. We have
\begin{equation}\label{eq:LP-general}
 \{F,G\}(\alpha):=\<\alpha,[\ud F,\ud G]\>,
\end{equation}
where $[\ud F,\ud G]$ is the Lie bracket of $\mathfrak{g}$ and $\<\cdot,\cdot\>$ is the pairing between the algebra and its dual space. To get an expression in coordinates of the Lie--Poisson bracket \eqref{eq:LP-general}, we can choose a basis $\{x^i\}_{i=1}^{\dim\mathfrak{g}}$ of $\mathfrak{g}$. In this coordinate system, we define the structure constants as $c^{ij}_kx^k:=[x^i,x^j]_k x^k$. We can identify the basis of $\mathfrak{g}$ with a coordinate system on its dual space $\mathfrak{g}^*$, and hence we find the well--known formula
\begin{equation}\label{eq:LP-coord}
 \{x^i,x^j\}=c^{ij}_kx^k.
\end{equation}
The concrete form of the bracket depends on the Lie algebra we are considering and in particular on whether it is finite or infinite dimensional. Let us consider the Lie algebra of vector fields on a two-dimensional domain $D$. To avoid technical difficulties, we assume that $D\equiv\R^2$ and that the fields are fastly decaying functions of $\vec{r}$. In some coordinates $\{x^i\}_{i=1}^2$, a vector field on $D$ can be written as $X(\vec{r})=\sum_{i=1}^2 X^i(\vec{r})\dev_i$; the components of the commutator are $[X,Y]^i(\vec{r})=\sum X^j(\vec{r})\dev_j Y^i(\vec{r})-Y^j(\vec{r})\dev_jX^i(\vec{r})$. This implies that the structure functions of the Lie algebra of vector fields $\mathfrak{X}$ must have the form $C^i_{jk}(\vec{r},\vec{r}',\vec{r}'')=\delta^i_j\delta(\vec{r}''-\vec{r})\dev_k\delta(\vec{r}'-\vec{r}'')-\delta^i_k\delta(\vec{r}'-\vec{r})\dev_j\delta(\vec{r}''-\vec{r}')$. The coordinates on $\mathfrak{X}^*$ are a set of functions $p_i(\vec{r})$ such that
\begin{equation*}
 \int_D p_i(\vec{r})v^i(\vec{r})\ud \vec{r}
\end{equation*}
behaves as a scalar under change of variables. Here, $v^i(\vec{r})$ are the components of a vector field. This means that $p_i(\vec{r})$ are densities of 1-forms. The Lie--Poisson bracket is linear in the coordinates and defined by the structure functions as
\begin{equation}\label{eq:LPbra}
 \{p_j(\vec{r}'),p_k(\vec{r}'')\}=\int_D C^i_{jk}(\vec{r},\vec{r}',\vec{r}'')p_i(\vec{r})\ud \vec{r}.
\end{equation}
The Poisson brackets is then defined by the density
\begin{equation}\label{eq:LPhPB_full}
 \{p_i(\vec{r}),p_j(\vec{r}')\}=\left(p_i(\vec{r})\frac{\dev}{\dev x^j}+p_j(\vec{r})\frac{\dev}{\dev x^i}+\frac{\dev p_i(\vec{r})}{\dev x^j}\right)\delta(\vec{r}-\vec{r}').
\end{equation}
This construction is due to Novikov \cite{N82}. The Poisson bracket \eqref{eq:LPhPB_full} is the Hamiltonian structure of EPDiff equation, namely the Euler--Poincar\'e equation associated with the diffeomorphisms group of a $n$-dimensional manifold. It has important applications in fluidodynamics -- it generalizes Camassa--Holm equation -- and even computer visions and imaging \cite{HM05}. In components, EPDiff equation has form
\begin{equation}\label{eq:EPDiff}
\frac{\dev m_i}{\dev t}+u^j\frac{\dev x^j}{m_i}+\frac{\dev u^j}{\dev x^i}m_j+\frac{\dev u^j}{\dev x^j}m_i=0,
\end{equation}
where $u^i$ are the components of the velocity field and we choose $m_i$'s as the conjugate momenta with respect to the Hamiltonian $H=\frac{1}{2}\int \vec{u}\cdot\vec{m}=\frac{1}{2}\int \eta^{ij}m_im_j$. Here $\eta^{ij}$ denotes the inverse of the metric on $D$. It is a straightforward to check that
$$
\frac{\dev m_i}{\dev t}=\{m_i,H\}
$$
is equivalent to \eqref{eq:EPDiff}, using the Poisson structure \eqref{eq:LPhPB_full} after relabelling $p_i$'s as $m_i$'s.

Let us perform the reduction of the Poisson structure from $\X^*$ to $\X'{}^*$, where with $\X'$ we denote the Lie algebra of the divergence free vector fields. We will reduce the structure functions of $\X$ and use them to define the reduced Lie--Poisson bracket. In dimension 2, as already discussed, the components of such vector fields can be written in terms of the scalar stream function: a more compact version of \eqref{eq:stream} is $X^i=\epsilon^{ij}\dev_j\psi$, with $\epsilon$ the two-dimensional Levi--Civita symbol.

In the infinite dimensional setting, the form of the commutator of two vector fields is
$$
[X,Y]^i(\vec{r})=\int_D C^i_{jk}(\vec{r}',\vec{r}'',\vec{r}')X^j(\vec{r}')Y^k(\vec{r}'')\ud\vec{r}'\ud\vec{r}''.
$$
Denoting with $\phi$ the stream function of the vector field $X$ and $\psi$ the stream function of $Y$, we rewrite the integral as
\begin{equation}
 [X,Y]^i(\vec{r})=\int_D \tilde{\mathcal{C}}^i(\vec{r}',\vec{r}'',\vec{r}')\phi(\vec{r}')\psi(\vec{r}'')\ud\vec{r}'\ud\vec{r}''.
\end{equation}
with
\begin{equation}
 \tilde{\mathcal{C}}^k(\vec{x},\vec{y},\vec{z})=\epsilon^{il}\epsilon^{jm}\frac{\dev}{\dev x^l}\frac{\dev }{\dev y^m}C^k_{ij}(\vec{x},\vec{y},\vec{z}).
\end{equation}
The commutator of two divergence free vector fields is a divergence free vector, hence we must have $[X,Y]^k=\epsilon^{kn}\dev_n\chi$ for a new stream function. After some manipulation we find the form for the structure function $\mathcal{C}$ of $\X'$ fulfilling $\chi(\vec{z})=\int\mathcal{C}(\vec{x},\vec{y},\vec{z})\phi(\vec{x})\psi(\vec{y})$. We have
\begin{equation}\label{eq:const-divfree}
 \mathcal{C}(\vec{x},\vec{y},\vec{z})=\epsilon^{lm}\frac{\dev}{\dev z^m}\delta(\vec{z}-\vec{x})\frac{\dev}{\dev z^l}\delta(\vec{z}-\vec{y}).
\end{equation}
To define the Lie--Poisson bracket we have to find the conjugate momentum of the stream function. From the Hamiltonian \eqref{eq:Ham-manip} we observe that the vorticity field $\omega$ plays that exact role. We can conclude that the Lie--Poisson bracket on $\X^*$ is
\begin{equation}\label{eq:LP-deriv}
\begin{split}
 \{\omega(\vec{x}),\omega(\vec{y})\}&=\int_D\epsilon^{lm}\frac{\dev}{\dev z^m}\delta(\vec{z}-\vec{x})\frac{\dev}{\dev z^l}\delta(\vec{z}-\vec{y})\omega(\vec{z})\ud\vec{z}\\
 &=\left(\frac{\dev\omega(\vec{x})}{\dev x^1}\frac{\dev}{\dev x^2}-\frac{\dev\omega(\vec{x})}{\dev x^2}\frac{\dev}{\dev x^1}\right)\delta(\vec{x}-\vec{y})
\end{split}
\end{equation}
as in \eqref{eq:PB-def}.
\section{Poisson Vertex Algebras and dispersive deformations of scalar Poisson brackets}
The language of Poisson Vertex Algebras is regarded as a very effective framework to study evolutionary Hamiltonian PDEs \cite{BdSK09}. In particular, it provides a fully algebraic formalism in which one can investigate the Hamiltonian structures, their symmetries and integrability, and so on. In this Section we will first briefly introduce the notion of multidimensional Poisson Vertex Algebra, of which the one associated to \eqref{eq:PB-def} is an example; then we will discuss the dispersive deformations of the bracket by direct computation.
\subsection{Multidimensional Poisson Vertex Algebra}
Let us consider the algebra $\A$ of the differential polynomials generated by $\omega$ and its derivatives. We will denote $\dev_{x^1}\omega=\omega_1$, $\dev_{x^2}\omega=\omega_2$, and similarly for higher order derivatives. Moreover, we assign a degree to the differential polynomials, by counting the order of the jets variables. For $f, g\in\A$, we have $\deg (fg)=\deg f +\deg g$, $\deg\omega=0$, $\deg\omega_I=\deg\mdev^I\omega=|I|$.
\begin{deff}
A ($D$-dimensional) Poisson Vertex Algebra (PVA) is a differential algebra $(\A,\{\dev_i\}_{i=1}^D)$ endowed with  $D$ commuting derivations and with a bilinear operation $\{\cdot_{\mlambda}\cdot\}\colon \A\otimes \A\to \R[\lambda_1,\ldots,\lambda_D]\otimes \A$ called the $\lambda$ \emph{bracket} satisfying the following set of properties:
\begin{enumerate}
\item $\{\dev_i f_{\mlambda} g\}=-\lambda_i\{f_{\mlambda} g\}$
\item $\{f_{\mlambda}\dev_i g\}=\left(\dev_\alpha+\lambda_i\right)\{f_{\mlambda} g\}$
\item $\{f_{\mlambda} gh\}=\{f_{\mlambda} g\}h+\{f_{\mlambda} h\}g$
\item $\{fg_{\mlambda} h\}=\{f_{\mlambda+\mdev}h\}g+\{g_{\mlambda+\mdev}h\}f$
\item $\{g_{\mlambda}f\}=-{}_{\to}\{f_{-\mlambda-\mdev}g\}$ (PVA-skewsymmetry)
\item $ \{f_{\mlambda}\{g_{\mmu}h\}\}-\{g_{\mmu}\{f_{\mlambda}h\}\}=\{\{f_{\mlambda}g\}_{\mlambda+\mmu}h\}$ (PVA-Jacobi identity).
\end{enumerate}
We use a multi-index notation  $\mlambda^I=\lambda_1^{i_1}\lambda_2^{i_2}\cdots\lambda_D^{i_D}$ for $I=(i_1,i_2,\ldots,i_D)$. The terms in the RHS of Property (4) are to be read, if $\{f_{\mlambda}h\}=\sum_I B(f,h)_I \mlambda^I$, as $\{f_{\mlambda+\mdev}h\}g=\sum_IB(f,h)_I(\mlambda+\mdev)^Ig=\sum_IB(f,h)_I(\lambda_1+\dev_1)^{i_1}\cdots(\lambda_D+\dev_D)^{i_D}g$. Similarly, the skewsymmetry property \emph{in extenso} is $\sum_I B(g,f)_I\mlambda^I=-\sum_I(-\mlambda-\mdev)^I B(f,g)_I$.
\end{deff}
The grading on $\A$ is extended to $\R[\mlambda]\otimes\A$ by imposing $\deg\mlambda^I=|I|$. The set of axioms for the PVA translates into a practical formula that gives the bracket between two elements of $\A$ in terms of the bracket between the generators $u^i$.
\begin{equation}\label{eq:master}
  \{f_{\mlambda}g\}=\sum_{\substack{i,j=1\ldots,N\\L,M\in\Z^D_{\geq 0}}}\frac{\dev g}{\dev u^j_M}(\mlambda+\mdev)^M\{u^i_{\mlambda+\mdev}u^j\}(-\mlambda-\mdev)^L\frac{\dev f}{\dev u^i_L}.
\end{equation} 

The relation between the notion of Poisson Vertex Algebra and the formal variational calculus is given by an isomorphism between the Poisson Vertex Algebras and the Poisson bracket on the space of local functionals \cite{MC15-1}. In particular, given the $\lambda$ bracket of a PVA we have that
\begin{equation}
 \left\{\int f,\int g\right\}=\int\{f_{\mlambda}g\}|_{\mlambda=0}
\end{equation}
and, given a Poisson bracket in the space of the local densities $\A$ 
$$\{u^\alpha(x),u^\beta(y)\}=\sum_I B^{\alpha\beta}_I(u(x);u_L(x))\mdev^I\delta(x-y),$$ we can define a $\lambda$ bracket
\begin{equation}\label{eq:defLambdaB}
 \{u^\alpha_{\mlambda}u^\beta\}=\sum_I B^{\beta\alpha}_S(u;u_L)\mlambda^S.
\end{equation}
Moreover, an evolutionary Hamiltonian PDEs of form $u_t=\{\int h,u\}$ is mapped to $u_t=\{h_{\mlambda}u\}|_{\mlambda=0}$.

In our case, $D=2$ and the differential algebra $\A$ is generated by the vorticity. The Poisson bracket \eqref{eq:PB-def} is translated by \eqref{eq:defLambdaB} to the $\lambda$ bracket
\begin{equation}\label{eq:PB-lambda}
 \{\omega_{\mlambda}\omega\}=\omega_1\lambda_2-\omega_2\lambda_1.
\end{equation}

One of the main advantages of the formalism of PVAs with respect to the standard technique is that the straightforward formula \eqref{eq:master} can be easily implemented to perform the explicit computations. In the next Paragraph we will study the deformations of the bracket \eqref{eq:PB-lambda} up to the second order.

\subsection{Deformations of the bracket}
Let us introduce the transformations
\begin{equation}\label{eq:Miuradef}
 \omega\mapsto \tilde\omega=\sum_{k=0}^\infty\epsilon^k F_k(\omega;\omega_I),
\end{equation}
on the space $\A$, where $F_k$ is a homogeneous differential polynomial of order $k$, and
\begin{equation*}
 \frac{\dev F_0(\omega)}{\dev \omega}\neq0.
\end{equation*}
The transformations \eqref{eq:Miuradef} form a group who is called the \emph{Miura group} \cite{DZ}. It can be regarded as the group of local diffeomorphisms on the space $\A$. The transformation of the 0 degree coordinates $\omega$ is then lifted to the higher degree jet variables $\omega_I$. An important subclass of Miura transformations, that plays a central role in the theory of the deformations, are the so-called \emph{second kind Miura deformations} \cite{lz11}, for which $F_0=\omega$.

\begin{deff}\label{defDefo}
 Given a $\lambda$ bracket $\{\cdot_{\mlambda}\cdot\}_0$, a $n$-th order infinitesimal \emph{compatible deformation} of it is a bracket
\begin{equation}
 \{\cdot_{\mlambda}\cdot\}=\{\cdot_{\mlambda}\cdot\}_0+\epsilon\{\cdot_{\mlambda}\cdot\}^\sim
\end{equation}
such that the properties (5) and (6) of the PVAs are satisfied up to order $\epsilon^n$. Moreover, the deformation is constituted by homogeneous terms
 \begin{equation}\label{defbrack}
 \{\cdot_{\mlambda}\cdot\}^{\sim}=\sum_{k=0}^n\epsilon^k\{\cdot_{\mlambda}\cdot\}_{k+1}.
 \end{equation}
in such a way that the bracket $\{\cdot_{\mlambda}\cdot\}$ is homogeneous after the assignment of degree $-1$ to the formal parameter $\epsilon$.
 In the case \eqref{eq:PB-lambda}, since the degree of the undeformed bracket is 2, we have that  $\deg \{\cdot_{\mlambda}\cdot\}_k$ is $k+2$.
\end{deff}
\begin{deff}\label{defTriv}
 A deformation of the $\lambda$ bracket \eqref{eq:PB-lambda} is said to be trivial if there exists an element $\phi$ of the Miura group such that
$$
\left\{\phi(\omega)_{\mlambda}\phi(\omega)\right\}_0=\phi\left(\{\omega_{\mlambda}\omega\}\right)+O(\epsilon^{n+2}).
$$
From Definition \ref{defDefo}, this implies that $\phi$ must be of second kind. 
\end{deff}
\begin{thm}
 The Poisson bracket of two-dimensional Euler's equation \eqref{eq:PB-def} does not admit nontrivial infinitesimal deformations up to the second order.
\end{thm}

The main result of this paper is obtained by direct computation of the deformations for the related $\lambda$ bracket \eqref{eq:PB-lambda}. The first order deformed bracket of \eqref{eq:PB-lambda} is a homogeneous $\lambda$ bracket of degree 3. Its general form, that depends on 36 parameters, is
\begin{equation}\label{eq:defo1-gen}
\begin{split}
 \{\omega_{\mlambda}\omega\}&=A^{abc}(\omega)\lambda_a\lambda_b\lambda_c+B^{ab,c}(\omega)\lambda_a\lambda_b\omega_c+C^{a,bc}(\omega)\lambda_a\omega_b\omega_c+\\
&+D^{a,bc}(\omega)\lambda_a\omega_{bc}+E^{abc}(\omega)\omega_{abc}+F^{a,bc}(\omega)\omega_a\omega_{bc}\\
&+G^{abc}(\omega)\omega_a\omega_b\omega_c.
\end{split}
\end{equation}
The commas in the indices are just a bookkeeping device to keep track of the different properties of symmetry, that are the ones corresponding to the symmetry of the expression. For instance, from the total symmetry in $(a,b,c)$ of $\lambda_a\lambda_b\lambda_c$ it follows that $A^{abc}$ must be totally symmetric in its upper indices. Imposing the skewsymmetry property to the bracket \eqref{eq:defo1-gen} we find that the parameters $A$'s, $C$'s, and $D$'s are left unconstrained, while the remaining ones must satisfy the conditions
\begin{align*}
 B^{ab,c}&=\frac{3}{2}A^{abc}{}'\\
 E^{abc}&=\frac{1}{12}\left(2D^{a,bc}+2D^{b,ca}+2D^{c,ab}-3A^{abc}{}'\right)\\
F^{a,bc}&=\frac{1}{4}\left(2C^{b,ca}+2C^{c,ab}+2D^{a,bc}{}'-3A^{abc}{}''\right)\\
G^{abc}&=\frac{1}{12}\left(2C^{a,bc}{}'+2C^{b,ca}{}'+2C^{c,ab}{}'-3A^{abc}{}'''\right).
\end{align*}
Imposing the fulfillment of the PVA--Jacobi identity generates a system of linear algebraic and differential equations for the remaining 16 parameters. The system is obtained by taking the first order in $\epsilon$ of Property (6). The result is a 5-th order differential polynomial in $\mlambda$, $\mmu$ and the jet variables $\omega_I$. The coefficients of each term of the polynomial
\begin{multline}\label{eq:compcond}
 \{\omega_{\mlambda}\{\omega_{\mmu}\omega\}_1\}_0-\{\omega_{\mmu}\{\omega_{\mlambda}\omega\}_1\}_0-\{\{\omega_{\mlambda}\omega\}_1{}_{\mlambda+\mmu}\omega\}_0\\
+\{\omega_{\mlambda}\{\omega_{\mmu}\omega\}_0\}_1-\{\omega_{\mmu}\{\omega_{\mlambda}\omega\}_0\}_1-\{\{\omega_{\mlambda}\omega\}_0{}_{\mlambda+\mmu}\omega\}_1=0
\end{multline}
give an overdetermined system of equations. In this case, however, it is enough to set equal to 0 the coefficients of the terms $\lambda^4\mu$, $\lambda^3\mu\dev\omega$, and $\lambda^2\mu(\dev\omega)^2$ to get that $A$'s, $C$'s, and $D$'s must all vanish. This means that there do not exist any first order deformation of \eqref{eq:PB-lambda}. This is always the case for a scalar bracket.

A second order deformation of the bracket depends has degree 4. In principle, it depends on 92 parameters, but after imposing the skewsymmetry property we are left with 36 free coefficients. A general fact occurring with scalar brackets is that the relevant coefficients are only the ones that multiply an odd number of $\lambda$'s. This means that for the bracket
\begin{equation}\label{eq:defo2-gen}
 \begin{split}
 \{\omega_{\mlambda}\omega\}&=A^{abc,d}(\omega)\lambda_a\lambda_b\lambda_c\omega_d+B^{ab,cd}(\omega)\lambda_a\lambda_b\omega_c\omega_d+C^{ab,cd}(\omega)\lambda_a\lambda_b\omega_{cd}\\
&+D^{a,bcd}(\omega)\lambda_a\omega_{bcd}+E^{a,b,cd}(\omega)\lambda_a\omega_b\omega_{cd}+F^{a,bcd}(\omega)\lambda_a\omega_b\omega_c\omega_d\\
&+G^{abcd}(\omega)\omega_{abcd}+H^{a,bcd}(\omega)\omega_a\omega{bcd}+I^{ab,cd}(\omega)\omega_a\omega_b\omega_{cd}\\
&+J^{ab,c}(\omega)\omega_{ab}\omega_{cd}+H^{abcd}\omega_a\omega_b\omega_c\omega_d.
\end{split}
\end{equation}
only the coefficients $A$'s, $D$'s, $E$'s, and $F$'s are independent, while the remaining can always be expressed in terms of them. Moreover, it is worthy noticing that there do not exist the terms in $\lambda^4$, since we cannot have a skewsymmetric scalar differential operator whose leading order is even. 

The huge set of equations we get after imposing the compatibility condition \eqref{eq:compcond} is composed both of algebraic and differential ones. In particular, it is possible to algebraically solve them for most of the free 36 parameters, resulting with six free parameters. After a change of coordinates
\begin{align*}
 C^{a,bcd}&\mapsto C^{a,bcd}-A^{abc,d}{}''-A^{abd,c}{}''-A^{acd,b}{}''\\
 D^{a,b,cd}&\mapsto D^{a,b,cd}-3A^{abc,d}{}'-3A^{abd,c}{}'-3A^{acd,b}{}'\\
 E^{a,bcd}&\mapsto E^{a,bcd}-A^{abc,d}-A^{abd,c}-A^{acd,b}\\
\end{align*}
they are $A^{112,1}$, $A^{122,1}$, $A^{222,1}$, $E^{1,112}$, $E^{1,122}$, and $E^{1,222}$.

Then, we consider the trivial deformations of \eqref{eq:PB-lambda}. We select the ones for which the Miura transformation is of degree two, in such a way that the resulting deformed bracket is of degree 4 as the one we are dealing with. Such a deformation has form
\begin{equation}
 \omega\mapsto f^{ab}(\omega)\omega_a\omega_b+g^{ab}(\omega)\omega_{ab}
\end{equation}
and depends on the six parameters $f$'s and $g$'s. We will compare the form of the six free parameters for a generic compatible deformation with the ones we get for a general trivial one, that in principle depends on six parameters, too. We can regard this latter set as a inhomogeneous system of algebraic and differential equations for $f$'s and $g$'s, whose explicit form is
\begin{align*}
 f^{11}&=-3 A^{112,1} & 2f^{12}&=-3A^{122,1} & f^{22}&=-A^{222,2} \\
 g^{11}-2f^{11}{}'&=2 E^{1,112} & g^{12}-2 f^{12}{}'&=E^{1,122} & 2g^{22}+f^{22}{}'&=-2E^{1,222}.
\end{align*}
This system is solvable for any values of $A$'s and $E$'s if and only if all the compatible deformations of the bracket, that are indeed parametrized by this six parameters, can be got from a Miura transformation of the coordinates; namely, if and only if there do not exist nontrivial compatible deformations of \eqref{eq:PB-lambda}. From the form of the system it is obvious that it can always be solved, since we can first trivially solve for $f$'s, then substitute their derivatives in the second set of equations and find $g$'s. The Theorem is proved.
\section{Concluding remarks}
In this paper we focussed on two aspects of the Hamiltonian structure of Euler's and Helmholtz's equations for a two-dimensional fluid. We have explicitly obtained the Poisson brackets as the reduction to the divergence free vector fields, that characterize the condition of ideal fluid, of the Lie--Poisson brackets associated to the full algebra of vector fields.

Moreover, we have established the triviality of the first and second order deformations of the bracket. It is a well known fact, proved by several authors \cite{G01,dGMS05,DZ}, that all the deformations of the one dimensional Poisson brackets of hydrodynamic type  are trivial. Using the language of PVAs, this means that the result holds true for brackets homogeneous of degree 1 and a differential algebra endowed with one derivation $\dev$, namely for fields depending on one space variable.

For the two-dimensional case it has been proved that the deformations at the first order are trivial \cite{MC15-1} while this is not the case for higher order deformations \cite{MC15-2}. Very recently it has been proved that the Poisson cohomology of multidimensional scalar brackets is extremely large \cite{CCS15}, hence there are plenty of deformations of such structures. The bracket \eqref{eq:PB-lambda} is at the crossroads of the aforementioned cases. It is a two-dimensional scalar bracket, namely it is defined for a scalar field of two variables, but it is not of hydrodynamic type, since its degree is 2. Nevertheless, it arises from the reduction of the Lie--Poisson bracket of hydrodynamic type for the Lie algebra $\X$, which is two-dimensional and two-component: the generators of $\A$ are the two fields $(p_1(x,y),p_2(x,y))$. In \cite{MC15-2} we have proved that the first order deformations of the Poisson bracket \eqref{eq:LPhPB_full} are trivial, while the second order ones span a two-dimensional space. Looking for higher order deformations of the bracket \eqref{eq:PB-lambda} requires much more computational effort. It may always be true that the odd order deformations are trivial, according to preliminary results for the third order. A more general method to compute the dimension of the cohomology groups of the bracket, as the one presented in \cite{CCS15}, maybe sufficient to prove the triviality of all the deformations, deserves further investigation.
\section*{Acknowledgements}

This work is partially supported by Italian National Institute of High Mathematics -- Group of Mathematical Physics Young Researcher Project 2014 ``Geometric and Analytic aspects of Integrable Systems'' and by PRIN 2010--11 Grant ``Geometric and analytic 
theory of Hamiltonian systems in finite and infinite dimensions'' of Italian Ministry of Universities and Researches. The author is grateful to the organizers of PMNP2015 for the opportunity to present his work and for the wonderful scientific environment they have established in Gallipoli.
\bibliography{biblio}

\end{document}